\documentclass[10pt,letterpaper]{article}
\usepackage[top=0.85in,left=2.75in,footskip=0.75in,marginparwidth=2in]{geometry}
\usepackage{soul}
\usepackage{graphicx}
\usepackage{hyperref}
\usepackage{placeins}
\usepackage{mathptmx}
\usepackage{anyfontsize}
\usepackage{t1enc}
\usepackage{lastpage,fancyhdr,graphicx}
\usepackage{epstopdf}
\usepackage{cite} 
\usepackage{multirow}
\usepackage{url}
\usepackage{tabularx}
\usepackage{placeins}
\usepackage{multicol,lipsum}
\usepackage{rotating}
\usepackage{natbib}
\usepackage[utf8]{inputenc}

\usepackage{cite}

\usepackage{nameref,hyperref}

\usepackage[right]{lineno}

\usepackage{microtype}
\DisableLigatures[f]{encoding = *, family = * }

\raggedright
\usepackage{setspace} 
\usepackage{changepage}
\usepackage{url}
\usepackage[aboveskip=1pt,labelfont=bf,labelsep=period,singlelinecheck=off]{caption}

\makeatletter
\renewcommand{\@biblabel}[1]{\quad#1.}
\makeatother

\usepackage{lastpage,fancyhdr,graphicx}
\usepackage{epstopdf}
\fancyheadoffset[L]{2.25in}
\fancyfootoffset[L]{2.25in}

\usepackage{color}

\definecolor{Gray}{gray}{.25}

\usepackage{graphicx}

\usepackage{sidecap}

\usepackage{wrapfig}
\usepackage[pscoord]{eso-pic}
\usepackage[fulladjust]{marginnote}
\reversemarginpar

\begin{document}
\vspace*{0.35in}

% title goes here:
\begin{flushleft}
{\Large
\textbf\newline{Automatic method for detection of
solar coronal width using extreme ultra-violet (EUV) radiation}
}
\newline
% authors go here:
\\
Najmeh Ahmadi\textsuperscript{1},
Shervin Parsi\textsuperscript{2}

\bigskip
\bf{1} Department of physics, Abdolrahman Sufi Razi Higher Educational Institute, 22 Bahman Hwy, Zanjan, 45371-38793, Iran,
\\
\bf{2} {Department of Physics, The Graduate Center, CUNY, New York, NY 10016 US}\\

\bigskip

\end{flushleft}
\justify
\section*{Abstract}
Solar corona, the last main layer of the atmosphere of the Sun, is detectable in the EUV and
X-ray. The corona is expanding into space up to millions of kilometers and is observable during the
eclipse. The temperature is increasing about millions of Kelvin. The investigation of this layer is
significant for solar physicists because it is dynamic and features. Active regions (AR) and solar
mass ejections (CMEs) are the important features in the solar corona. In this research, the solar limb and coronal width
is studied from full-disk images at $284 \AA$  taken by SOHO/EIT during eleven-year
period (2000-2010). Next, using image processing methods and by applying region growing
function, the corona is segmented and extracted from images in different angles. The radial velocities of CMEs are extracted.

\textbf{Key words:} Sun: corona, Sun: extreme ultra violet (EUV), technique: segmentation.

% now start line numbers

% the * after section prevents numbering
\section{Introduction}
The study of the Sun is developing through the combination of observational data and numerical MHD simulation (see e.g. \cite{ganjali2019resonant}). Wave propagation in the magnetic flux tubes is one of the most fundamental problems of MHD theory to describe the physics behind the several solar phenomena. In this regard, a comprehensive research about the wave propagation inside and outside the magnetic flux tubes has been studied by \cite{Esmaeili_2016, Esmaeili_2017, Esmaeili_2015}. On the other hand, the solar surface is covered by small-scale features during the solar cycles. Statistical properties of these features are important subjects of solar physics \citep{Javaherian_2014, Arish_2016, Javaherian_2017, Yousefzadeh_2015_a, Yousefzadeh_2015_b, Honarbakhsh_2016}.

\cite{Rozelot_2015} analyzed radius dependence on wavelength (from X-ray to radio waves)
using the data sets published in the literature. \cite{Lefebvre_2004} also developed a new method of detecting features at the solar limb, such as active regions (sunspots and faculae) or supergranulation. \cite{dam_2000} review the scientific goals linked to the diameter measurements. \cite{Verbeeck_2013} introduced a set of segmentation procedures (known as the SPoCA-suite) that allows one to retrieve AR and CH properties on EUV images taken by SOHO-EIT, STEREO-EUVI, PROBA2-SWAP, and SDO-AIA. 

\cite{Selhorst_2019} analyzed the lower limit of the polar brightening observed at 100 and 230 GHz by ALMA, during its Science Verification period, 2015 December 16-20. They  found that the average polar intensity is higher than the disk intensity at 100 and 230 GHz.

\cite{Kosovichev_2018} presented results of a new analysis of twenty one years of helioseismology data from  SoHO  and SDO, which  resolve previous uncertainties and compare variations of the seismic radius in two solar cycles. 

\cite{Rozelot_2018} investigated  that the Their new look on  modern measurements of
the Sun’s global changes from 1996 to 2017 gives a new way for peering into the solar
interior and play an important role in the implementation of the solar cycles.

\cite{Rozelot_2011} provided a new perspective on the controversy which followed measurements made in Princeton (USA) in the mid-1960s. They pointed out how false ideas or inexact past measurements may contribute to the advancement of new physical concepts. \cite{Rozelot_2016} also studied the Evolution of the subsurface meridional flows obtained from the 5-years of the SDO/HMI observations during Solar Cycle 24. \cite{Lefebvre_2007} studied the solar surface radius variations from SOHO during solar cycle 23 (1996-2005) to put constraints on the radius changes. \cite{Kilcik_2009} also investigated that the  eclipse data show significant variations of the solar radius observed over time scales shorter than one year. d

This paper is organized as follows, our proposed method  is introduced in Section \ref{sec2}. In Section \ref{sec3}, the results of implementing our proposed method are presented.  The main conclusions is also presented in Section \ref{sec4}.

\section{Method}\label{sec2}
Using the methods of images processing and applying the function of region growing for the coronal segmentation, we address the area of this region of the sun's atmosphere at different line of sides. To this end, after applying a circular mask on the disk, we segment the remaining regions by applying the growth function of the region; and according to the nature of the problem, one or more points are selected as seed for the growth of the region. First, we select an arbitrary pixel in the region as the seed cell. The next step is to apply threshold intensity, which is very important to select this intensity. Here, we test some intensities; and the threshold intensity for the above function will give us better images. We choose the same intensity for all the images. The higher the threshold intensity, the more pixels are selected. On the other hand, by selecting a low threshold, some parts of the regions may be lost. However, all the pixels in the atmosphere are specified by selecting a proper threshold. In the next step, after selecting the proper threshold, the four seed point neighbors are subjected to threshold testing to join or not join the seed.

\begin{figure*}[ht]
\centerline{\includegraphics[width=1\textwidth,clip=]{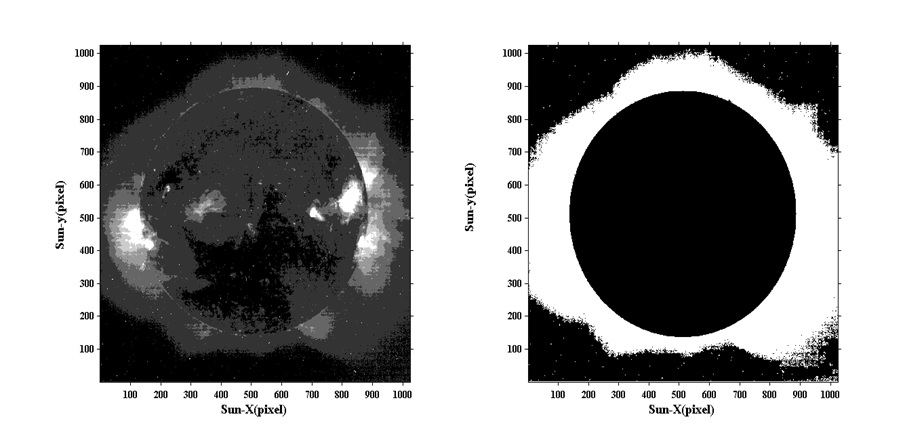}}
\caption{Left: Original image at 284 \AA wavelength taken by SOHO satellite on 7 o’clock, January 10, 2005; right: image of the corona area after applying the growth function of the region separated from the background.}\label{fig1}      
\end{figure*}

This trend occurs for all pixels in the region, until the regions join and create the desired region with the desired threshold intensity. The image segmentation is performed based on intense changes in the intensity values such as image’s edges. The result of applying function of region growing along with its cut image is shown for an image, Fig \ref{fig1}. 

\begin{figure*}[ht]
\centerline{\includegraphics[width=1\textwidth,clip=]{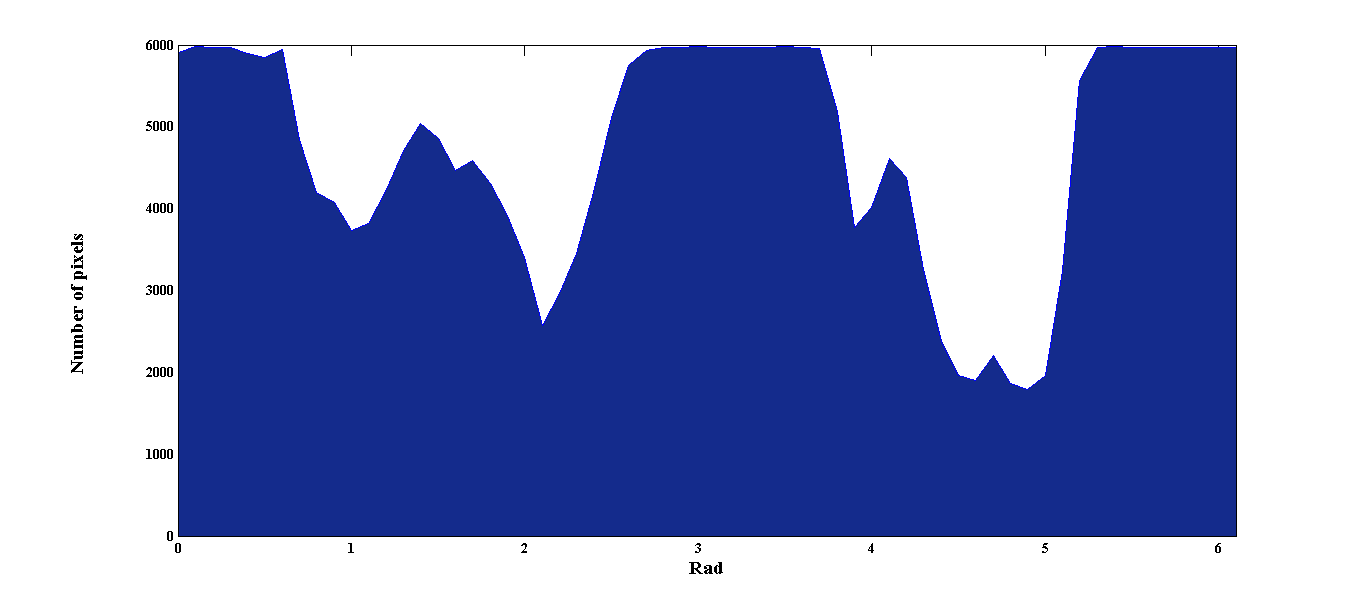}}
\caption{The number of pixels in different line of sides.}\label{fig2}      
\end{figure*}

The obtained region is investigated at different angles. The chart of the number of pixels at different angles is shown in Fig. \ref{fig2}. As observed in this figure, and by analyzing the consecutive data series in this wavelength, the number of pixels in the region is plotted in terms of angle. It has peaks and valleys; the peaks represent the fact that the corona has the highest atmosphere at the peaks; and in the valleys, the corona atmosphere is at the lowest level, indicating a periodic principle in the solar atmosphere chart.
In this research, by collecting data\footnote{\url{http://sdac.virtualsolar.org/cgi/cartui}} for an 11-year cycle, we did the same and obtain a chart for corona width for each day. We could also plot the mean corona width for each year and extract the maximum and minimum values of each year. In the following, the average corona width was plotted for the entire 11-year cycle. We also recorded data about active regions and CMEs\footnote{\url{www.lmsal.com/hek/hek/isolsearch.html}}, designed to guide sun researchers, in order to investigate the statistical data on the sun. The diagram of the distribution of solar latitude and longitude of the active regions, the mean histogram of the active regions’ area, and the diagram of the CMEs’ histogram are plotted based on radial velocity for all years. Ultimately, the maximum and minimum values have been investigated in all of these charts, and the relationship between the active regions, CMEs and solar latitude and longitude has been analyzed.

\section{Results}\label{sec3}
In this research, we have calculated the corona width over an 11-year solar cycle using ultraviolet images at a wavelength of 284 \AA. To this end, SOHO satellite data (AIT) have been employed. To advance our goals, we have used programs written in MATLAB environment. The output of the software is in charts based on the numbers in different angles (intervals of 1 rad.). As observed in the diagrams, the extracted widths have maximums and minimums, indicating the periodic changes of the corona over the 11-year cycle. In this chapter, first the average corona width in the 11 years is investigated. Then, in the next step, the physical parameters of the active regions, such as solar latitude and longitude, and their area are represented in the charts for the 11-year cycle. Finally, the radial velocity distribution of the CMEs is presented.

\subsection*{Coronal Width}

\begin{figure*}[ht]
\centerline{\includegraphics[width=1\textwidth,clip=]{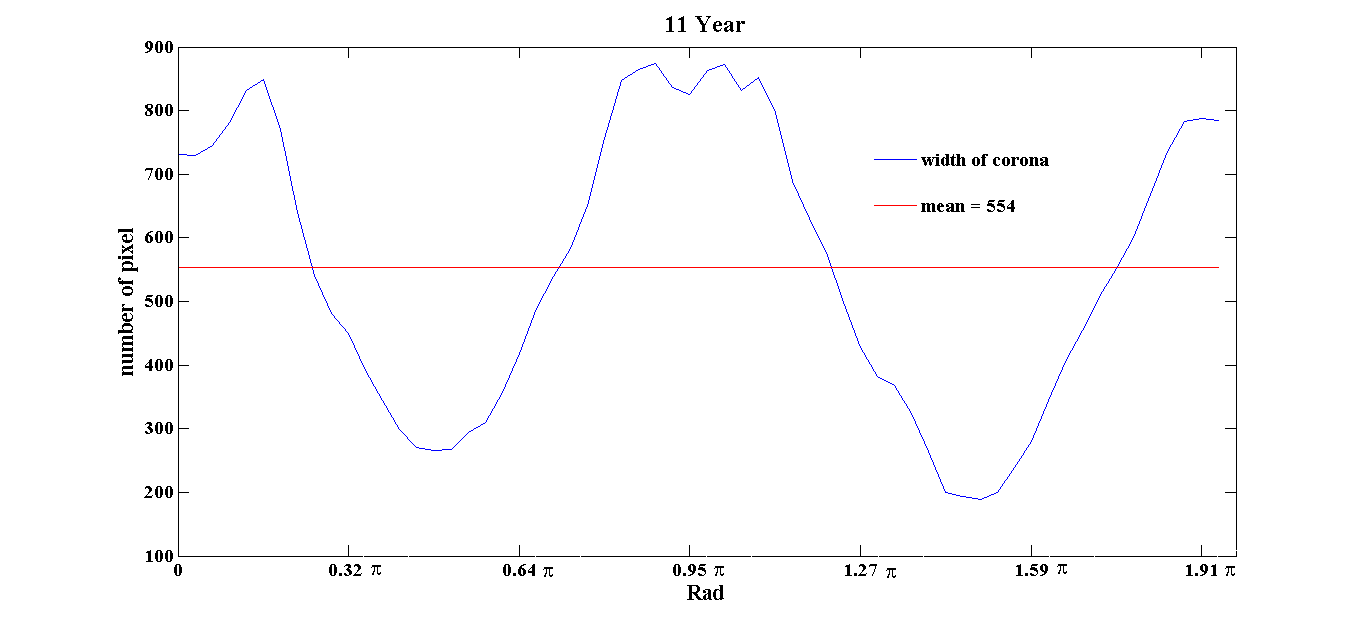}}
\caption{Distribution of pixels by angle for an 11-year cycle.}\label{fig5}      
\end{figure*}

As observed in Fig. \ref{fig5}, the chart has some maximum and minimum values. It can be claimed that the corona has the highest atmosphere in the peaks and its lowest value in the valleys. The average corona width is shown about 554 pixels. This chart has two maximum values of about 5.2 to 2.3 radians and two minimum values in the values of 5.1 to 5.4.
\begin{table}[]
\centering
\caption{Average, minimum and maximum values (in mega meters) for coronal width in an 11-year cycle.}
\label{tab1}
\begin{tabular}{|l|l|l|l|}
min & max & average & year  \\
1878  & 4078   & 2846    & 2003 \\
488   & 4145   & 2064    & 2004 \\
473   & 3763   & 1842    & 2005 \\
213   & 2573   & 1126    & 2006 \\
0     & 2294   & 406     & 2007 \\
336   & 2119   & 1175    & 2008 \\
359   & 1868   & 1098    & 2009 \\
522   & 2604   & 1446    & 2010 \\
204   & 2988   & 1282    & 2011 \\
183   & 3078   & 1536    & 2012 \\
295   & 3625   & 1609    & 2013
\end{tabular}
\end{table}
The average corona width in 2003 has been more than other years. As expected, in this year, the solar activity has been also higher; and in 2007, it is less than the other years, and as expected, the solar activity is low.

\subsection{Distribution of Active Regions (in Area)}
\begin{figure*}[ht]
\centerline{\includegraphics[width=1\textwidth,clip=]{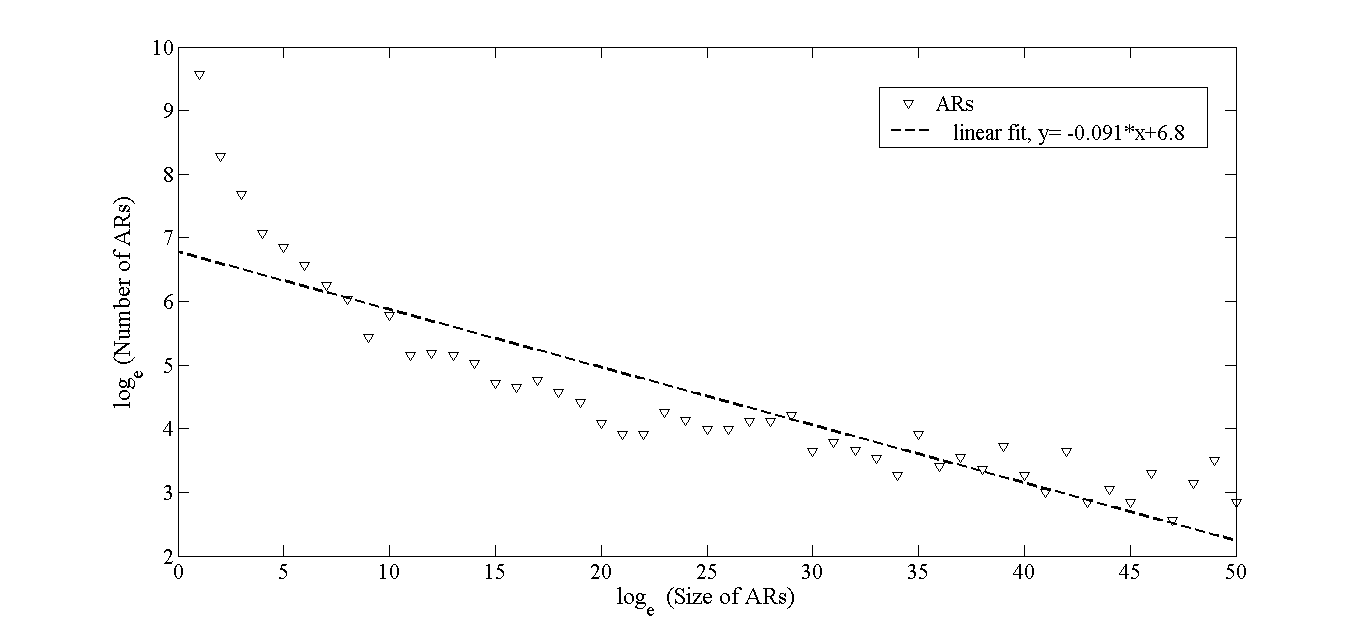}}
\caption{The distribution of active regions (in area) for an 11-year cycle.}\label{fig6}      
\end{figure*}

In the following, the area of active regions over a solar activity cycle from 2000 to 2010 is represented using a series of data. In Fig. \ref{fig6}, the distribution of area of the active solar regions in a solar cycle has been represented.
The maximum area of the active regions is about 20 mega meters. On the chart, the line $y = ax+b$ is fitted in the logarithmic representation with the slope of $a = -0.091$. The average area of active regions is about 743 square mega meters. The minimum and maximum values are 21 and 9900 square mega meters, respectively.

\subsection{Radial Velocity}
The radial velocity of coronal mass ejections (CMEs) has been extracted by means of various recorded data in a solar cycle from 2000 to 2010. The radial velocity distribution is represented in Chart 4.5. The maximum radial velocity of the coronal mass ejections is at a velocity of $220 km/s$ and the velocity range has been recorded at the minimum of 99 and maximum of 999.
 \begin{figure*}[ht]
\centerline{\includegraphics[width=1\textwidth,clip=]{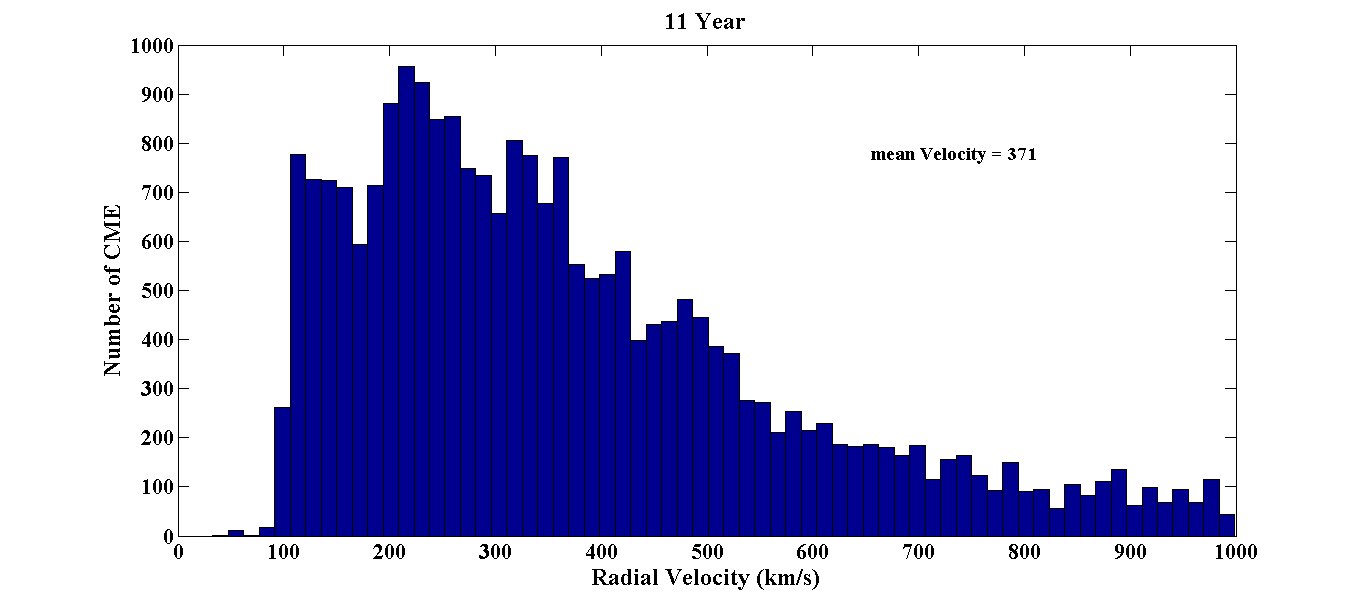}}
\caption{The radial velocity in terms of the number of CMEs for an 11-year cycle.}\label{fig7}      
\end{figure*}   

\section{Conclusion}\label{sec4}
In this study, for measuring the area of this region of the solar atmosphere at different angles, we used the image processing methods and applied the growth function of the region in order to segment the corona. Then, after applying a circular mask on the pill, we segmented the remaining regions by applying the growth function of the region and selected appropriate threshold intensity, and this trend was occurred for the entire pixels and corona width was plotted. Moreover, changes in solar latitude and longitude, radial velocity and active regions distribution were obtained.

The corona width over a cycle has significant changes and varies from 150 to 850 mega meters. Therefore, a layer with specific corona boundaries cannot be provided at all times. The number of active regions in an 11-year solar cycle has a significant equivalent periodic cycle. The solar cycle is obtained from counting the number of sunspots known as the butterfly chart. To extract possible alternatives, the obtained chart for the number of active regions in a cycle requires more analysis for extracting possible cycles. Currently, the existing data are available at most for solar cycles 23 and 24, and the debate may be evolved in the future with the completion of data. To this end, more data are needed to be analyzed for several cycles, so that the period of the number of active regions can be extracted in terms of time.

\nolinenumbers

%This is where your bibliography is generated. Make sure that your .bib file is actually called library.bib
\bibliography{library}

%This defines the bibliographies style. Search online for a list of available styles.
\bibliographystyle{rusnat}
\setcitestyle{authoryear}
\end{document}